# Convergent beam electron-diffraction investigation of lattice mismatch and static disorder in GaAs/GaAs$_{1-x}$N$_x$ intercalated GaAs/GaAs$_{1-x}$N$_x$:H heterostructures

S. Frabboni[1,2(a)], , V. Grillo[2], G. C. Gazzadi[2], R. Balboni[3], R. Trotta[4], A. Polimeni[4], M. Capizzi[4], F. Martelli[5,6], S. Rubini[5], G. Guzzinati[1,7], F. Glas[8]

[1]Dipartimento di Fisica, Università di Modena e Reggio Emilia, Via G. Campi 213/a, 41125 Modena, Italy
[2]CNR-Istituto di Nanoscienze-S3 Modena, Via G. Campi 213/a, 41125 Modena, Italy
[3]CNR-IMM Bologna, Via P. Gobetti 101, 40129 Bologna, Italy
[4]Dipartimento di Fisica, Sapienza Università di Roma, P.le A. Moro 2, 00185 Roma, Italy
[5]IOM-CNR, Area Science Park, S.S. 14, Km. 163.5, 34012 Trieste, Italy
[6]IMM-CNR, Via del Fosso del cavaliere 100, 00133 Roma, Italy
[7]EMAT, University of Antwerp, B-2020 Antwerp, Belgium
[8]CNRS-Laboratoire de Photonique et de Nanostructures, Route de Nozay, 91460 Marcoussis, France

Hydrogen incorporation in diluted nitride semiconductors dramatically modifies the electronic and structural properties of the crystal through the creation of nitrogen-hydrogen complexes. We report a convergent beam electron-diffraction characterization of diluted nitride semiconductor-heterostructures patterned at a sub-micron scale and selectively exposed to hydrogen. We present a method to determine separately perpendicular mismatch and static disorder in pristine and hydrogenated heterostructures. The roles of chemical composition and strain on static disorder have been separately assessed.

Diluted Ga(AsN) is a quite interesting semiconductor system because of the large mismatch and strong deformation of local lattice parameter due to N insertion in the host lattice.[1] Moreover, Ga(AsN) hydrogenation strongly influences the electronic properties of the system. Indeed, H forms stable N-$H_n$ (n>1) complexes with N, thus wiping-out N effects on both band gap and lattice parameter and modifying in a controllable way most physical properties of the material.[2] In GaAs/GaAs$_{1-x}$N$_x$ heterostructures, static disorder and perpendicular mismatch, defined as $m_\perp$= ($a_\perp$-$a_{sub}$)/ $a_{sub}$, where $a_{sub}$ and $a_\perp$ are the lattice parameters of the substrate and of the nitride along the growth direction, respectively, have been studied by means of high resolution X-ray diffraction and Rutherford Back-Scattering (RBS) channeling techniques.[3] These measurements have been then correlated with the H-induced changes of the electronic properties.[3] Recently, quantum confinement has been reported in GaAs$_{1-x}$N$_x$/GaAs$_{1-x}$N$_x$:H heterostructures.[4]

It is well known that a detailed and quantitative structural characterization of patterned structures requires high spatial resolution techniques. In Si/SiGe heterostructures, Convergent Beam Electron-Diffraction (CBED) and Large-Angle CBED (LACBED) have been proved to be able to measure $m_\perp$ and static disorder with a spatial resolution at the 10 nm level.[5] In particular, CBED methods have been applied also to the study of the strain field in electronic devices where both in-depth and lateral spatial resolution were necessary.[6-9,10-12]

The application of these high spatial resolution techniques to patterned physical systems whose properties are controlled by clusters of point defects is the issue we tackle here. The twofold aim of the present work is indeed: First to develop a high spatial resolution method for the study of mismatch and static disorder in heterostructures of sub-micron GaAs$_{1-x}$N$_x$/GaAs wires intercalated with micrometric GaAs$_{1-x}$N$_x$:H/GaAs strips. Second, to single out the role of chemical composition and strain on electron diffraction by exploiting the remarkable fact that these two factors can be separately controlled in the quasi-quaternary $GaAs_{1-x}N_x$:H.[2]

CBED and LACBED techniques have been applied to a patterned 120 nm-thick $GaAs_{1-x}N_x$ layer ($x$=2.2%) grown at 500 °C by molecular beam epitaxy on top of a GaAs buffer deposited at 600 °C on a (001) GaAs substrate. A 50 nm-thick film of titanium was deposited on the sample surface, previously capped with a 25 nm thick GaAs film and patterned by electron-beam lithography, to obtain 500nm-wide Ti wires separated by 5 μm. The patterned sample was hydrogenated at 300 °C by a low-energy (100 eV) ion beam with a $3 \times 10^{18}$ cm$^{-2}$ dose of impinging H atoms.[3] The cross-section sample for TEM (125 nm thick) was prepared with the lift-out method by using the 30 keV Ga$^+$ beam of the Focused Ion Beam (FIB) of a FEI Strata 235 Dual Beam equipment. A final polishing at 5 keV was performed with the same ion beam. This method allows the preparation of cross-sections with controlled and almost uniform thickness and with the necessary high spatial resolution selectivity of the thinning site and well fits measurements of $m_\perp$ and static disorder. In the TEM bright field image shown in Figure 1, crosses indicate the points where CBED patterns have been recorded, double arrows indicate the $GaAs_{1-x}N_x/GaAs_{1-x}N_x$:H interfaces normal to the sample growth plane and single arrows mark the interface with the GaAs substrate.

Due to the complexity of the strain field produced by the relaxation phenomena in thin TEM sample with multiple interfaces, finite element simulations have been performed[13] in order to find the variations of the most relevant lattice parameters describing the strain field. It has been found that the stress along the thinning direction is almost fully relaxed at approximately 30 nm from the lateral and in-depth interfaces. This finding simplifies the $m_\perp$ measurement reducing the number of lattice constants to be extracted from a single CBED pattern from 6 to 3 and thus allowing to find a unique solution for the strain tensor.[14] In addition, in this particular stress configuration the $m_\perp$ values can be retrieved by dividing the perpendicular mismatch measured in the thinned sample by a proper factor (1-ν),[15] where ν=0.31 is the isotropic GaAsN Poisson ratio.[16]

CBED experiments have been performed then by tilting the sample along the [230] zone axis and using an electron beam energy of 200 keV. The central disks of the CBED patterns, where the strain sensitive High Order Laue Zone (HOLZ) lines are formed,[17] are shown in Figure 2. The intersections between (-1,1,11) and (-1,1,-11) HOLZ lines, which are particularly sensitive to the variations of the lattice parameter, are indicated by arrows. In order to extract the strain data, we adopt a well-known procedure that permits to best fit the processed experimental patterns with quasi-kinematically simulated patterns.[7] The contrast of the HOLZ lines is then enhanced by filtering the Hough transform of each HOLZ line pattern and the lattice parameters of the strained layer are determined. In Figure 2a is shown the CBED pattern originating from the GaAs substrate taken as a reference for the determination of the effective acceleration voltage (199.85±0.05 keV).

In GaAsN, $m_\perp$ is negative ($m_\perp^{GaAsN}$ = -0.015±0.001, namely, the lattice parameter along the growth direction is smaller than that of the substrate, Figure 2b), slightly positive, instead, in GaAsN:H ($m_\perp^{GaAsN:H}$ = +0.005±0.001, Figure 2c), in agreement with previous measurements in unpatterned heterostructures.[3]

While mismatch is detected through the angular shifts of high-angle diffracted beams, static disorder reduces the coherent Bragg diffracted intensity.[18] A quantitative description of static disorder can be obtained by the exponential increase of the extinction distance $\xi_g$ –due to the increase of the mean square deviation of atoms from their lattice positions caused by the substitutional impurities[18]

$$\xi_g^* = \frac{\Omega_c \times \pi \times \cos(\theta_B)}{\lambda \times F_g \times e^{-2\pi^2 \times <u_x^2> \times g^2}} = \frac{\Omega_c \times \pi \times \cos(\theta_B)}{\lambda \times F_g \times e^{-M_{SD} \times g^2}} = \xi_g \times e^{M_{SD} \times g^2} \,, \qquad (1)$$

Here, $\Omega_c$ is the volume of the unit cell, $\theta_B$ is the Bragg angle, $\lambda$ is the de Broglie electron wavelength, $F_g$ is the structure factor calculated without taking into account atomic displacements, **g** is the operating reciprocal lattice vector, and $<u_x^2>$ is the mean square displacement parallel to **g**. Finally, $M_{SD}$ is the Debye-Waller factor accounting for static displacement in the alloy. Equation (1) clearly indicates that the effects on electron diffracted intensities are more easily observable for diffraction at large-angles. A simple relationship between the extinction distance and the integrated intensity of the electron diffracted beam has been given by Vainshtein[19]

$$\frac{I_g}{I_0} = \left(\frac{\pi}{\xi_g}\right)^2 \times t \times K(t,\xi_g) \times L , \qquad (2)$$

where $I_0$ is the incident electron beam intensity, $t$ is the sample thickness, $K$ is the oscillating dynamical factor, and $L$ is the Lorentz factor –in a perfect crystal set equal to one.[19]

In Figure 3, the integrated intensity of the electron diffracted beam, as calculated for different values of $t$, is shown by a blue line through full squares for **g**=(660) at 200 keV in GaAs ($\xi_g^{GaAs}$ =331nm),[20] and for $I_0$=1 (in GaAs$_{0.988}$N$_{0.022}$, $\xi_g^{GaAsN:H}$ =336 nm for the same value of **g**, neglecting static disorder effects).[20] In the same plot, the integrated intensities of the diffracted electron beams, as calculated by approximating for $(\pi t/\xi_g) <2$ the oscillating dynamical factor in Eq. 2 with a Gaussian function:

$$K(t,\xi_g) = e^{-\left(\frac{\pi \times t}{\sqrt{3} \times \xi_g}\right)^2} , \qquad (3)$$

are shown for increasing values of static disorder, namely, increasing $M_{SD}$. The integrated intensity of this reflection clearly scales with $M_{SD}$. Then, in a cross-section sample, the ratio between the integrated intensities of the electron diffracted beams, measured in GaAsN or GaAsN:H and normalized to GaAs and to the proper thickness ratio, would allow the determination of $M_{SD}$ in the different parts of the intercalated heterostructures. The thickness of the sample cross-section can be determined by a scanning electron microscopy (SEM) inspection of the lamella produced by FIB and checked by CBED measurements at low-angle reflection, such as the **g**=(220), whose rocking curve is almost insensitive to $M_{SD}$. In the present case, the sample thickness was almost uniform along the direction parallel to the surface, with a small in-depth gradient –125 nm in GaAsN and 150 nm in the GaAs buffer layer, respectively. Therefore, in the present sample prepared by FIB the analysis of static disorder has been performed around the maximum of the integrated intensity curve shown in Figure. 3, namely, where the ratio between the integrated intensities is weakly dependent on $t$ and well represented by the Vainshtein approximation of the oscillating dynamical factor given by Eq. (3). Thus,

$$\frac{I_g^{GaAsN:H} / I_g^{GaAs}}{I_g^{GaAsN} / I_g^{GaAs}} = \left(\frac{\xi_g^{GaAsN} \times e^{M_{SD}}}{\xi_g^{GaAsN:H}}\right)^2 \times \frac{t^{GaAsN:H}}{t^{GaAsN}} \times \left(\frac{e^{-\left(\frac{\pi \times t^{GaAsN:H}}{\sqrt{3} \times \xi_g^{GaAs}}\right)^2}}{e^{-\left(\frac{\pi \times t^{GaAsN}}{\sqrt{3} \times \xi_g^{GaAsN} \times e^{M_{SD}}}\right)^2}}\right), \quad (4)$$

The extinction distances $\xi_g^{GaAsN:H}$ and $\xi_g^{GaAsN}$ have been calculated without taking into account static atomic displacements, which are considered separately in the $e^{M_{SD}}$ term, and neglecting the effect of H.[20] Once the intensities ratios are measured, Eq. (4) can be solved with respect to $M_{SD}$. Although diffuse scattering is not taken into account in the present calculations, it can be reasonably assumed that the contribution to the experimental intensity patterns of both diffuse and inelastic scattering is minimized by a linear background subtraction, at least for high-angle reflections.[21]

A quantitative analysis of the intensity ratio defined in Eq. (4) can be done by Large Angle CBED.[22] In this case, different regions of the sample contribute to different parts of the same electron diffraction pattern, thus permitting an accurate comparison between diffracted intensities originating from different area of the sample.[5] LACBED experiments have been performed by tilting the sample near the [441] zone axis, where the high-angle reflection **g**= (660) (g=15 nm$^{-1}$, $\theta_B$=18.83mrad at 200keV in GaAs) can be brought into Bragg position and does not show strong dynamical interactions with other reflections. The corresponding Bragg contour, which is normal to the GaAs/GaAsN and GaAs/GaAsN:H interfaces, can be then quantitatively analyzed. Moreover, this reflection is not sensitive to atomic displacements, **r**, along the [001] and [1-10] directions normal to the free surfaces of the bulk sample and of the TEM cross-section, respectively –**g**•**r**=0, indeed. Unwanted effects, e.g., due to a stress relaxation induced by the specimen preparation, are thus minimized.

LACBED patterns recorded in Bright Field acquisition mode (BFLACBED), with the (660) deficiency Bragg contour at the center of the convergence disk, and Dark Field acquisition mode (DFLACBED), with the (660) excess Bragg contour also in the center of the convergence disk, are shown in Figure 4. These patterns have been taken in pristine [(a) and (b)] and hydrogenated [(c) and (d)] area of the samples, respectively. Figure 4e shows the depth profile of the integrated intensity of the **g**=(660) reflection originating from the pristine heterostructures (Figure 4b) and the hydrogenated one (Figure 4d), normalized to a mean value measured in the GaAs substrate, after a linear background subtraction. The reduction of diffracted intensity in the nitride layer can then be ascribed to static disorder induced by the presence of N atoms in the GaAs host lattice and by the chemical effect of the (As-N) substitution on the scattering factor. The average reduction of intensity is 0.78 and only 0.93 in the GaAsN and GaAsN:H layers, respectively. Static disorder is then partially compensated in the hydrogenated part of the heterostructures, in agreement with a slight overcompensation of the perpendicular mismatch found by CBED. We can then reasonably assume that GaAsN:H is a fully disordered alloy with an average perfect lattice whose scattering factor differs from GaAs only for chemical composition. The reduction of intensity between GaAsN and GaAsN:H would then entirely be due to the static disorder in GaAsN and result in a root mean square displacement, $u = \sqrt{\langle u_x^2 \rangle} = (0.058 \pm 0.002)$ Å.

This result has been obtained assuming a single exponential damping factor with an interpolated static Debye-Waller factor. More precisely, using $u_{III} = 0.0376$ Å and $u_V = 0.0646$ Å,[23] and imposing that

$$(f_{III} + f_V) \times e^{-2 \times \pi^2 \times g^2 \times u_{GaAsN}^2} = \left( f_{III} \times e^{-2 \times \pi^2 \times g^2 \times u_{III}^2} + f_V \times e^{-2 \times \pi^2 \times g^2 \times u_V^2} \right), \quad (5)$$

where $f_{III}$ and $f_V$ are the atomic scattering factors of group III and group V atoms, respectively, we find $u=0.0539$Å, with a reasonable agreement between theory and experiment. Near the bottom GaAsN:H/GaAs interface, a lack of compensation is observed, most likely because of an incomplete hydrogenation of the nitride layer or to an interface effect. This feature and the related strain field will be investigated elsewhere.

In conclusion, CBED-LACBED can determine the perpendicular mismatch and static disorder in diluted nitride heterostructures patterned to form sub-micron as-grown GaAsN wires intercalated with H irradiated GaAsN strips. In particular, these electron diffraction methods enable an experimental determination of the static atomic displacement due to chemical substitution of As with N. The present experiments improve our quantitative understanding of the role of static disorder on electron diffraction and open interesting perspectives for the analysis by electron microscopy methods of patterned systems whose physical properties are controlled by clusters of point defects, in particular, by N-$H_n$ clusters that play a fundamental role in diluted nitride heterostructures.

**AKNOWLEDGEMENTS**

The contribution of M. Francardi e A. Gerardino on sample processing is kindly acknowledged.

[a)] Author to whom correspondence should be addressed
Electronic mail: stefano.frabboni@unimore.it

**FIGURE CAPTIONS**

Figure 1. (a) TEM Bright field image of a $GaAs_{1-x}N_x$ /$GaAs_{1-x}N_x$:H heterostructure on top of a GaAs substrate. Points where CBED patterns have been recorded are marked by crosses, interfaces are marked by arrows.

Figure 2. HOLZ lines in the central disk of CBED patterns taken in: GaAs buffer (a), GaAsN (b), and GaAsN:H (c). Intersections between (-1111) and (-11-11) HOLZ lines, particularly sensitive to lattice parameter variations, are indicated by arrows.

Figure 3. The integrated intensity of the **g**=(660) reflection of the electron beam is shown as a function of the thickness of the sample by a blue through squares. The same quantity, as calculated in the Vainshtein approximation, see text, is shown for different values of the static Debye-Waller factor $M_{SD}$.

Figure 4. LACBED and DFLACBED with the (660) Bragg contour crossing the pristine [(a) and (b)] and hydrogenated [(c) and (d)] parts of the intercalated heterostructures. (e) Comparison between the integrated diffracted intensities originating from (b) and (d) showing a reduction of diffracted intensity in GaAsN with respect to GaAs that can be justified by the presence of substitutional static disorder in the alloy. Static disorder in the GaAsN layer is partially compensated by hydrogenation. A lack of compensation is found starting at ~30 nm from the bottom, deeper GaAsN:H/GaAs interface.

**FIGURE 1**

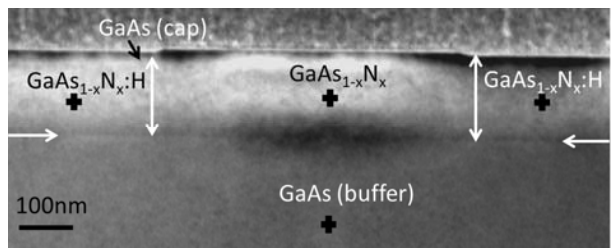

**FIGURE 2**

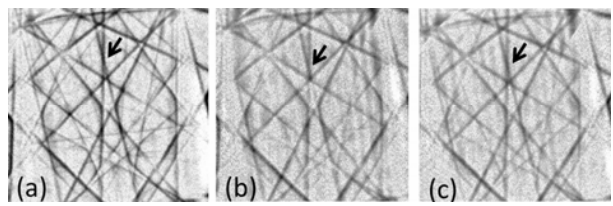

**FIGURE 3**

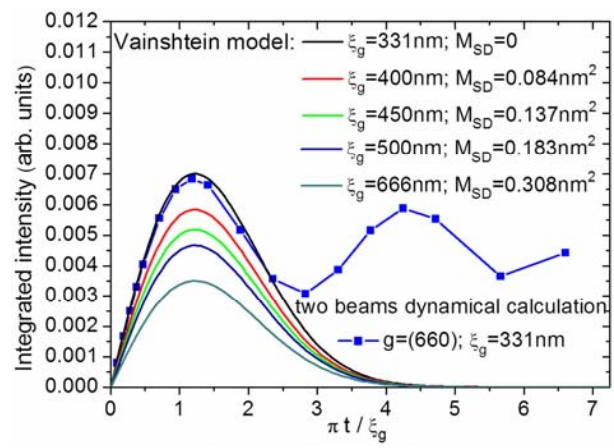

**FIGURE 4**

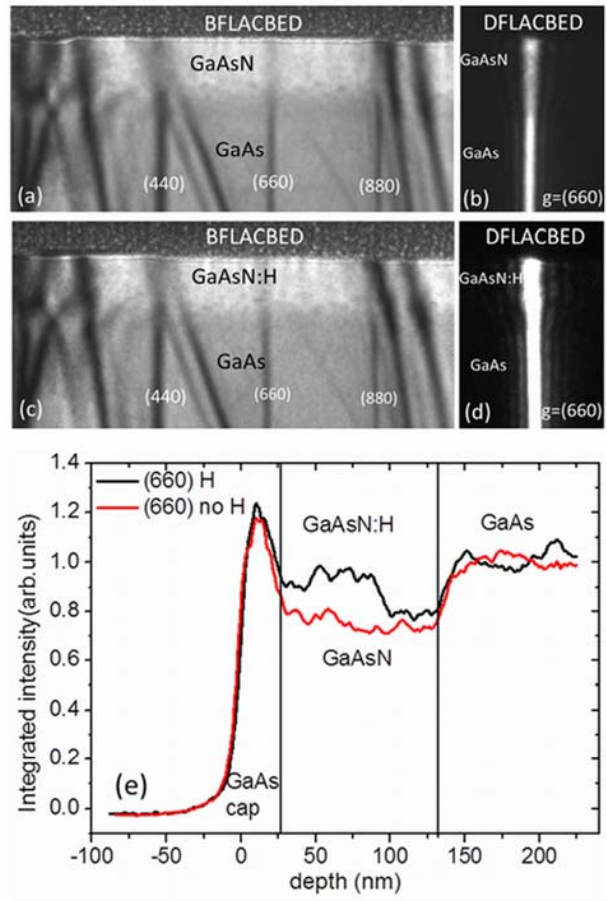

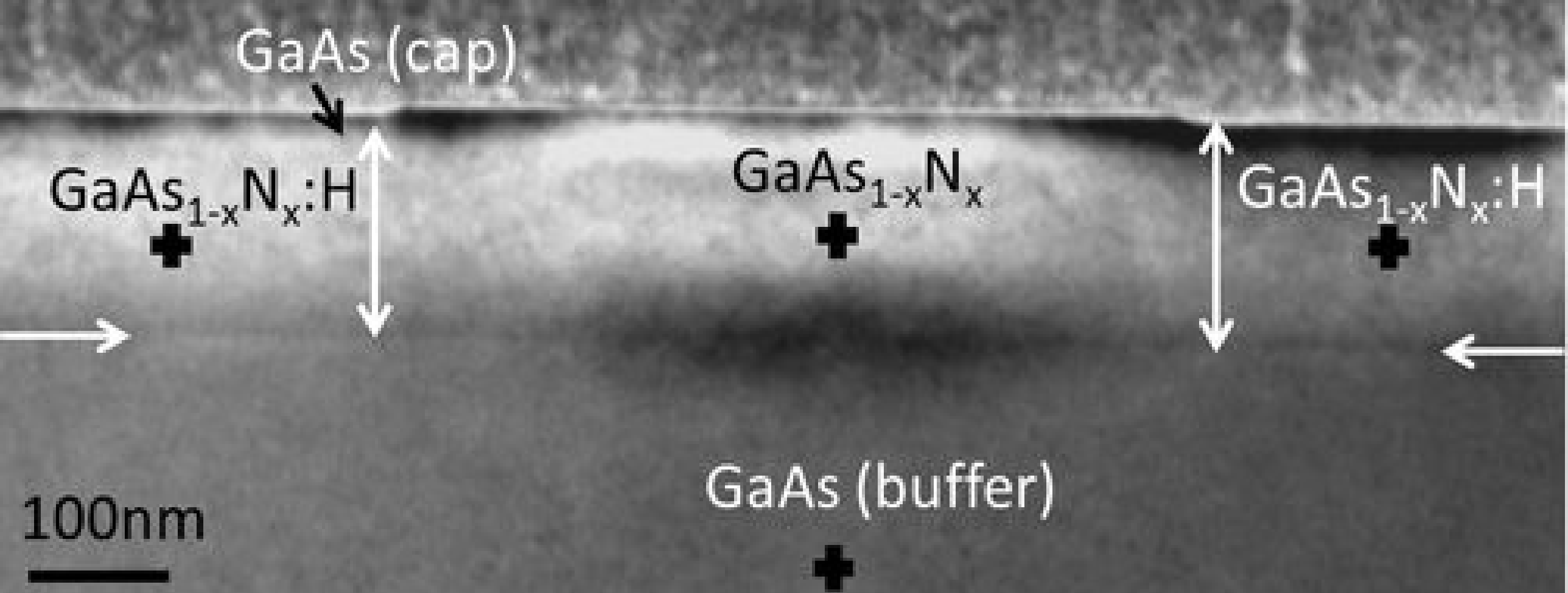

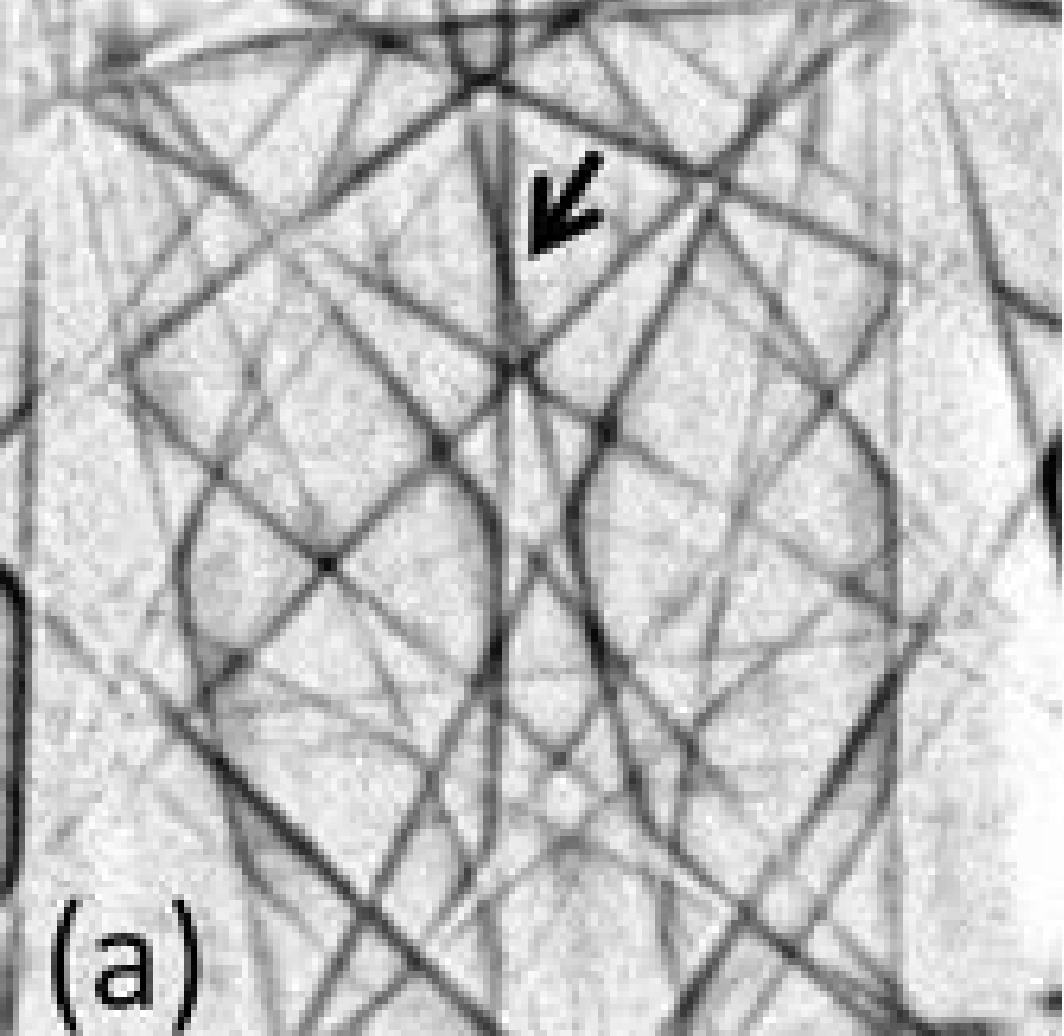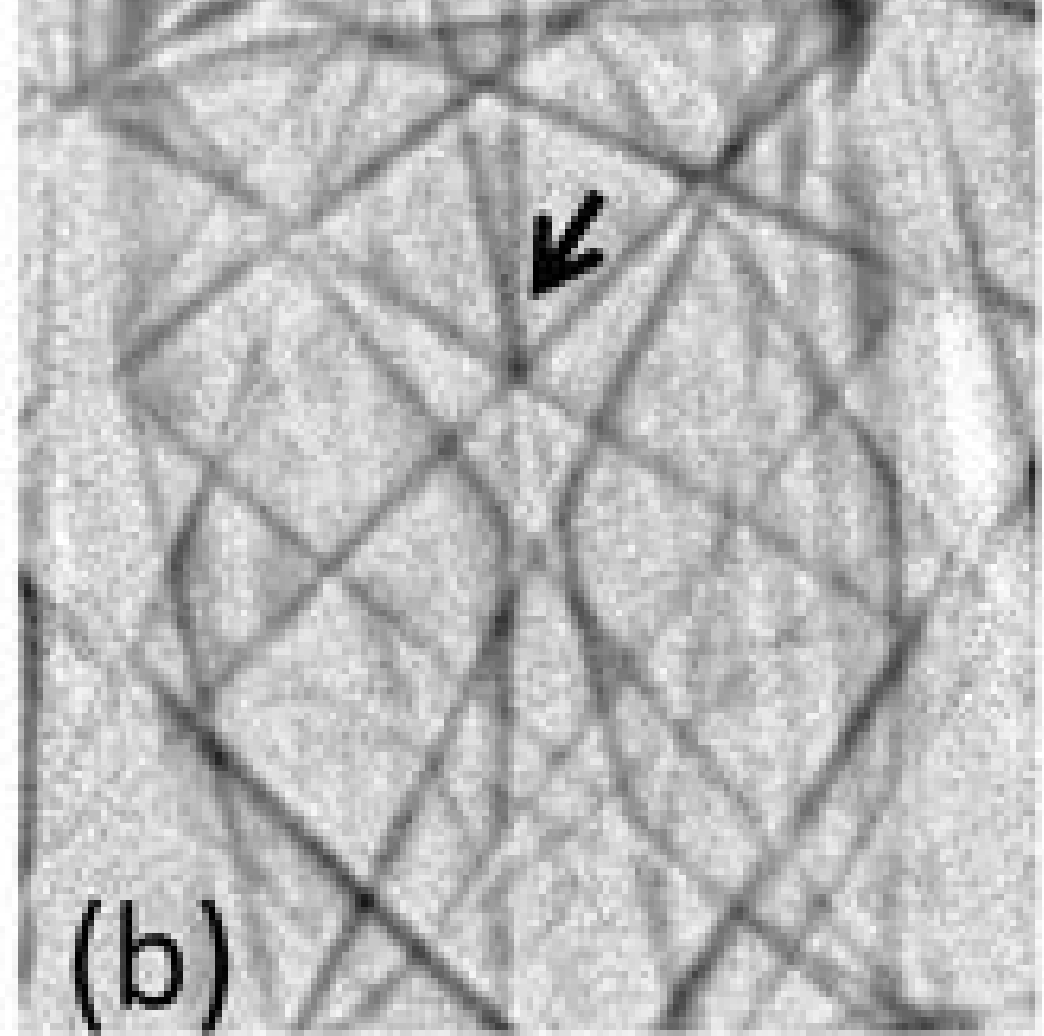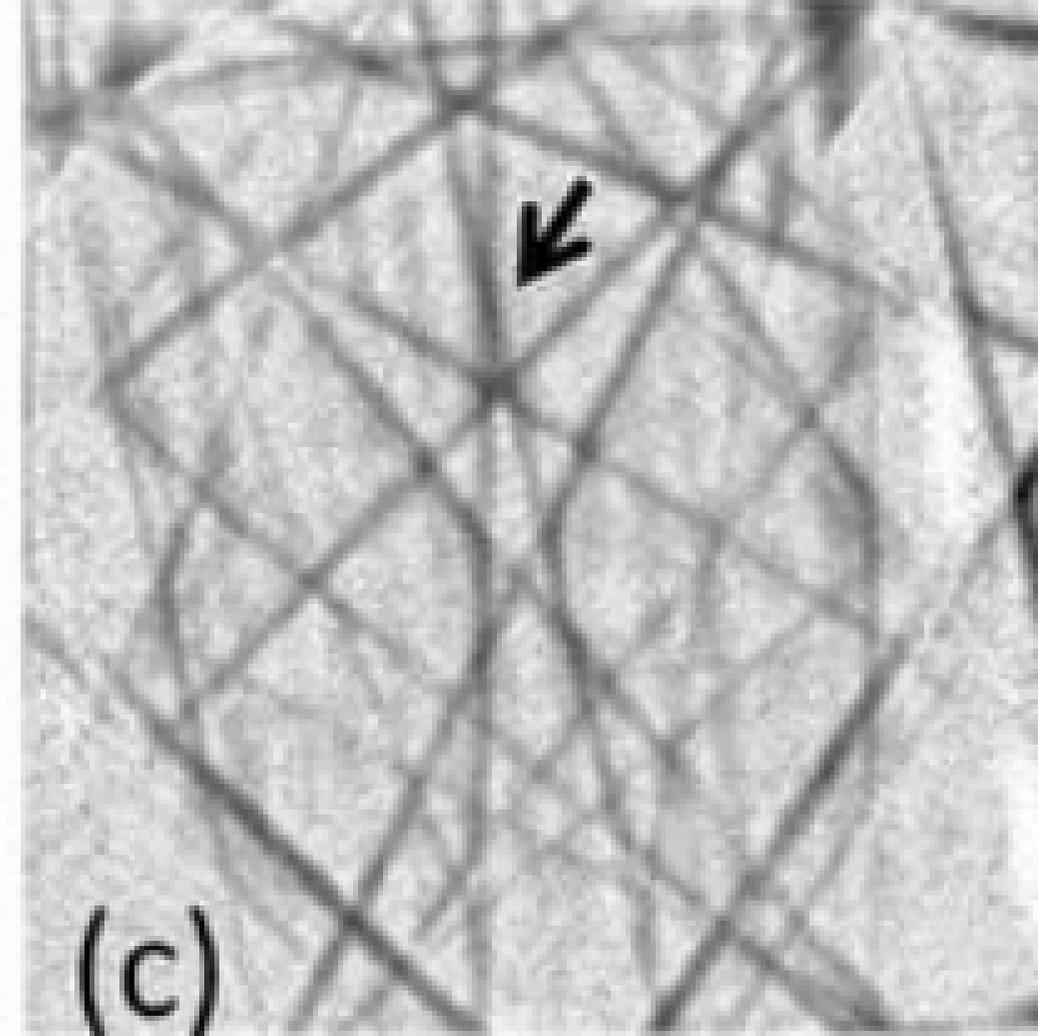

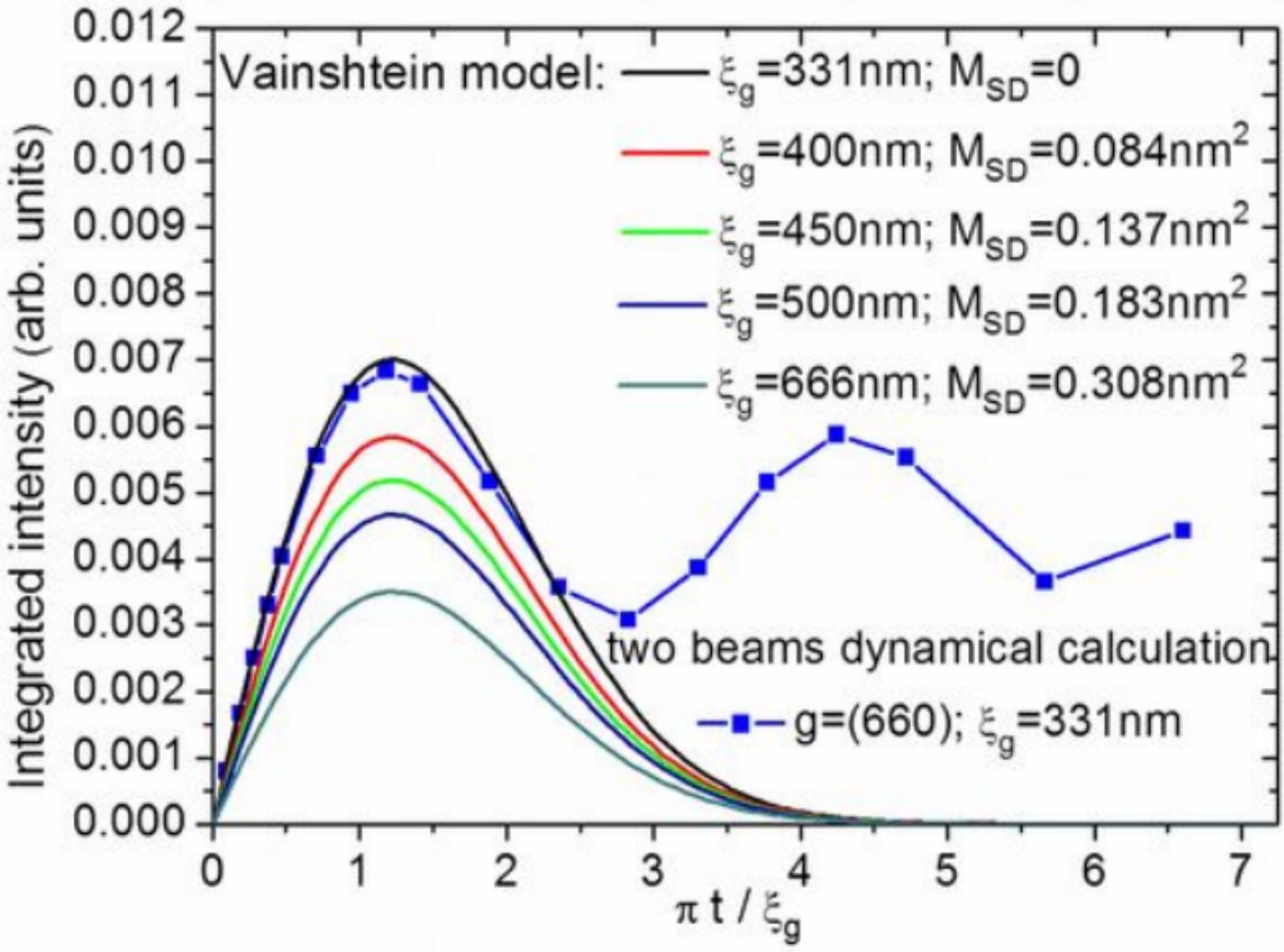

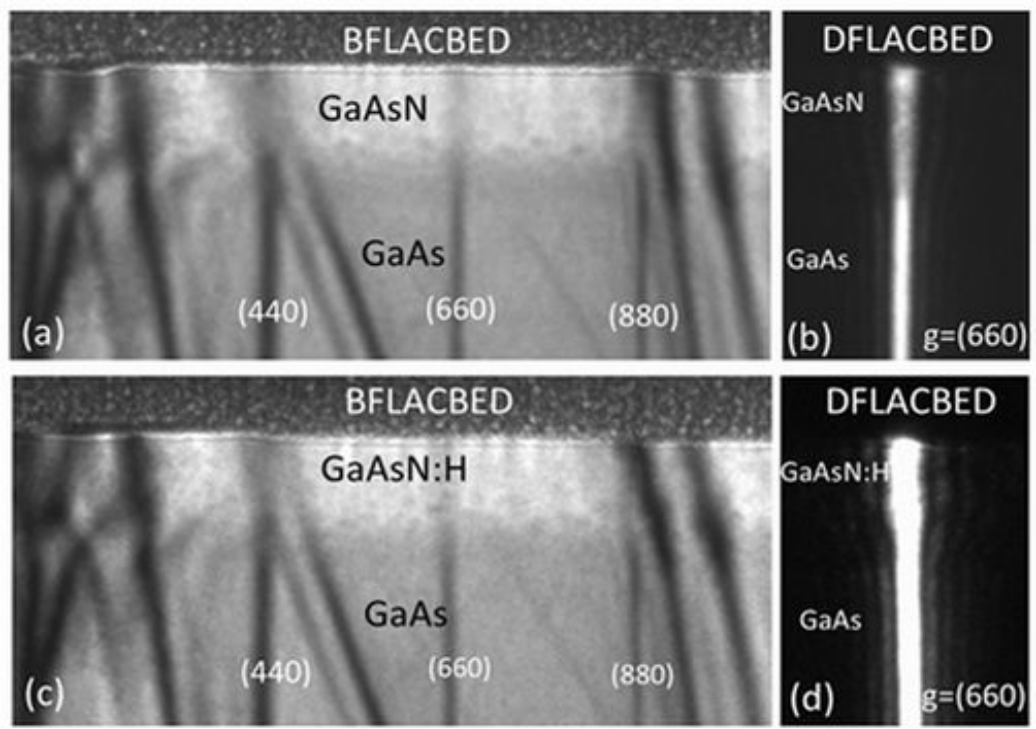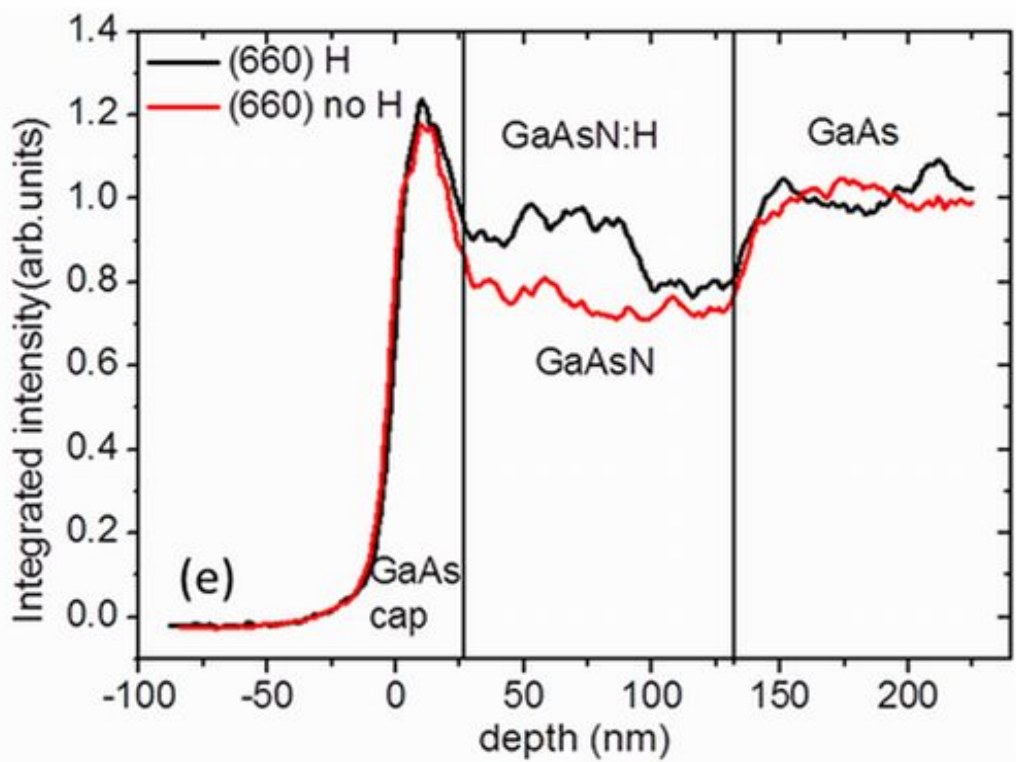